%% file: main.tex
\documentclass[10pt,conference]{IEEEtran}

\input{settings}

\begin{document}

\title{Snapshot Metrics Are Not Enough: Analyzing Software Repositories with Longitudinal Metrics}

\input{authors}

\maketitle

\begin{abstract}

Software metrics capture information about software development processes and products.
These metrics support decision-making, e.g., in team management or dependency selection.
However, existing metrics tools measure only a snapshot of a software project.
Little attention has been given to enabling engineers to reason about metric trends over time---longitudinal metrics that give insight about process, not just product.
In this work, we present \tool (PRocess MEtrics), a tool for computing and visualizing process metrics. 
The currently-supported metrics include productivity, issue density, issue spoilage, and bus factor. 
We illustrate the value of longitudinal data and conclude with a research agenda.
The tool's demo video can be watched at \href{https://youtu.be/YigEHy3_JCo}{\url{https://youtu.be/YigEHy3_JCo}}.
The source code can be found at \href{https://github.com/SoftwareSystemsLaboratory/prime}{github.com/SoftwareSystemsLaboratory/prime}.
\end{abstract}


\section{Introduction}
\label{sec:introduction}

An effective software engineering process is correlated with high software quality~\cite{sommerville2015software}.
Measurements of software processes give engineers insight into software quality~\cite{fenton_software_2014}.
Software metrics characterize the software engineering process (e.g., time to fix a defect) and the engineered product (e.g., cyclomatic complexity). 
Using software metrics, engineers and managers improve products and assess the risks of external software dependencies.


In our survey of existing tools for software metrics, we found that typical metrics tools provide aggregate metrics, or 'snapshot metrics,' rather than longitudinal metrics (\cref{sec:background}).
They capture the current state of the software product and perhaps some process metrics (e.g., age of current issues).
While a snapshot can be useful---for example, it can quickly reveal if a project has no test suite---it does not provide a full picture of the longitudinal evolution of a software project.
\emph{We conjecture that engineers will make different decisions when presented with snapshot metrics compared to considering the trends in a project's metrics over time (\cref{sec:demonstration})}.


To evaluate a development process, one needs to measure the history of the code.
The classic reference on software metrics~\cite{fenton_software_2014} establishes that measurement needs to be related to a time range and scale for a meaningful longitudinal assessment of software quality~\cite{fenton_software_2014}.
Tools that measure quality need to calculate both direct measurements and derived calculations at consistent intervals to evaluate the process properly.
Historical trends can be visualized and utilized to quantify software engineering decisions by collecting and reporting process metrics at consistent intervals.

To support our investigation of this research question, we present \tool~\cite{prime} (PRocess MEtrics): an open-source tool that enables engineers and researchers to analyze software projects with longitudinal metrics.
\tool uses a modular Extract-Transform-Load pipeline architecture for ease of adoption and extension.
\tool currently supports the following metrics: code size, productivity, bus factor, issue count, issue spoilage, and issue density.

We close by proposing three studies facilitated by \tool: 
(1) exploring engineers' use of longitudinal metrics when assessing their products;
(2) exploring their use of longitudinal metrics during dependency selection;
and
(3) analyzing the software supply chain to identify potential weak links. 


\section{Background and Related Work}
\label{sec:background}


Process metrics are critical for improving software quality as agile repositories may eventually become more established and require regular maintenance.
Although numerous efforts have focused on mining open-source repositories, the current support for  process metrics---and visualizing them longitudinally---is mixed.
In our survey of related efforts, we encountered various tool types, including scorecards, frameworks, dashboards, and platform monitors.

\emph{Scorecards} assign a risk score for open source projects to assess security risks and project health.
The \textsc{ossf/scorecard}~\cite{scorecard}
However, they are computed as a snapshot metric cannot easily express longitudinal effects.

\emph{Frameworks} simplify the mining software repository (MSR) development process.
These are typically domain system languages (DSL) and libraries that researchers and engineers integrate into their tools.
\textsc{ishepard/pydriller}~\cite{spadini_pydriller_2018} and the \textsc{Boa}~\cite{dyer_boa_2013} DSL meet this criteria.
These frameworks allow for MSR tools to be developed for the analysis of VCS, but they are not MSR tools.

\emph{Dashboards}
are built into online VCS platforms designed to provide visualizations of repository and issue tracker trends.
GitHub Insights~\cite{github_insights} and GitLab Insights~\cite{gitlab_insights} provide longitudinal metrics for hosted projects.
However, these tools provide limited insights when it comes to process metrics.

\emph{Platform monitors}
are third party analysis tools that compute metrics for hosted packages.
NPM~\cite{npm} implements the NPM Search~\cite{npms.io} monitor for JavaScript packages.
While the GoReportCard~\cite{goreportcard} is a monitor for Go projects hosted on GitHub.
The GoReportCard tracks code metrics, while NPM Search tracks process metrics regarding issue trackers.

Aside from dashboards, these tools compute process metrics as snapshots and do not make longitudinal and trends visualization easy for users.


\section{Architecture}
\label{sec:architecture}

\begin{figure}[h]
\centering
\includegraphics[width=0.48\textwidth]{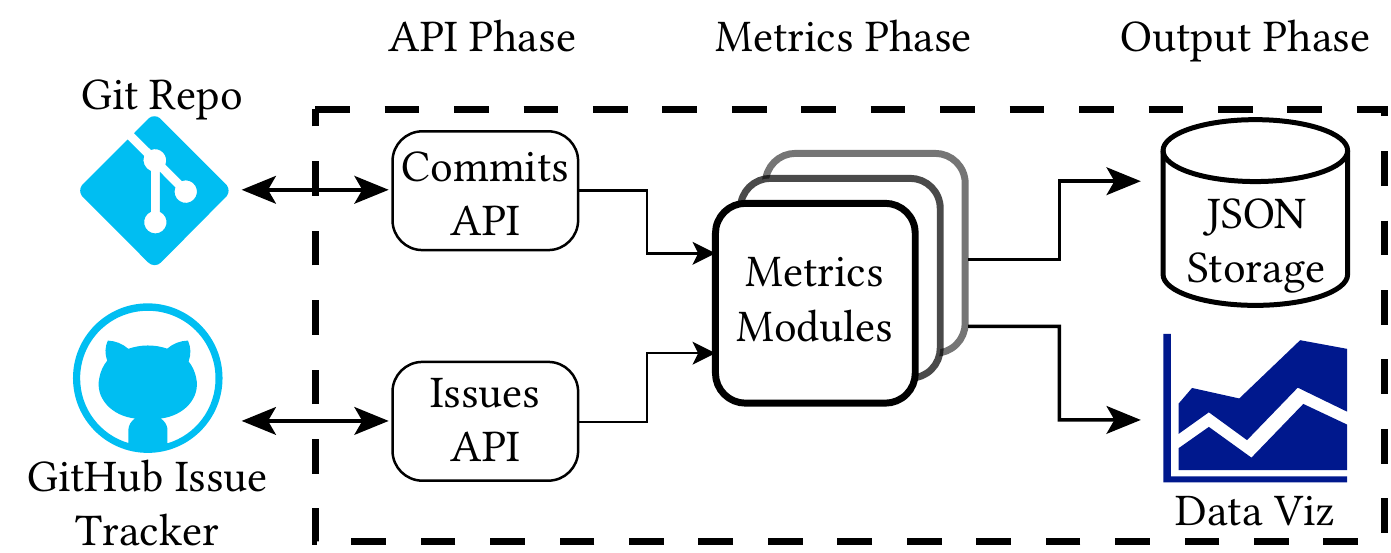}
\caption{System architecture of \tool.}
\label{fig:sys-arch}
\end{figure}

\tool is architected as an Extract, Transform, Load (ETL) pipeline that utilizes modular components to compute and visualize metrics, as shown in Figure\ref{fig:sys-arch}.

The ETL phases of the pipeline are each module or collection of modules.
In addition, the extraction and transformation stages of the pipeline store data in text-encoded JSON files.
By storing measurements in a file rather than in memory during pipeline execution, \tool can be integrated with existing tools and pipelines.

\tool extracts base measurements from a project's version control system (VCS) and issue tracker during the \textbf{API Phase}.
Here, using \textsc{cloc}~\cite{aldanial_2021_5760077} and \textsc{sloccount}~\cite{sloccount}, \tool measures each commit of a repository and measures the size of the repository in lines of code (LOC), thousands of lines of code (KLOC), and the size difference between each sequential commit as the delta thousands of lines of code (DKLOC).
\tool also extracts issue report metadata by utilizing the REST API of a repository's host issue tracker.

\tool transforms the extracted base measurements into derived metrics during its \textbf{Metrics Phase}.
At the moment, \tool can compute the following metrics issue spoilage, issue/defect density, productivity, and bus factor.
Each metric module takes in a text-encoded JSON file containing both or either the commits or issue base measurements.

After both the API and Metrics phases, data is loaded into either text-encoded JSON files or visualized with MatPlotLib~\cite{hunter_matplotlib_2007} in the \textbf{Output Phase}.
\tool can export the JSON and visualization files to integrate with other pipelines.
Additionally, the visualizations can be customized using style sheets and parameters for the major components, thereby allowing engineering teams to implement style standards for their visualizations.

This architecture allows engineers to download and install individual \tool modules to measure specific attributes of their software without waiting for measurements of unwanted attributes.
Furthermore, each phase of the pipeline is configurable, reducing the time engineering teams need to post-process the data to match their specific needs.

\begin{figure*}[ht!] 
    \centering
  \subfloat[\label{sub:redis-id}]{%
       \includegraphics[width=0.45\linewidth]{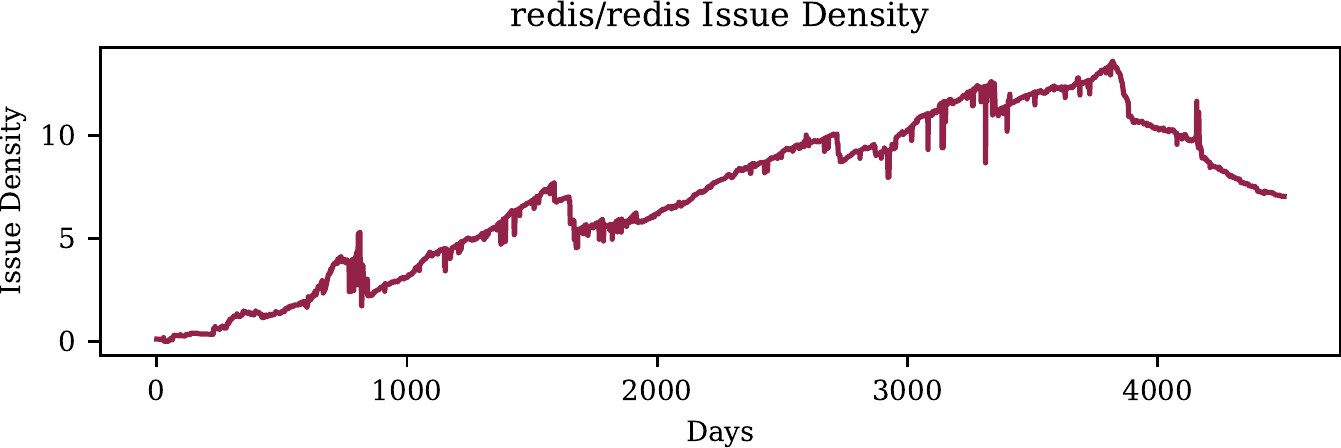}}
    \hfill
 \subfloat[\label{sub:pulp-or-id}]{%
      \includegraphics[width=0.45\linewidth]{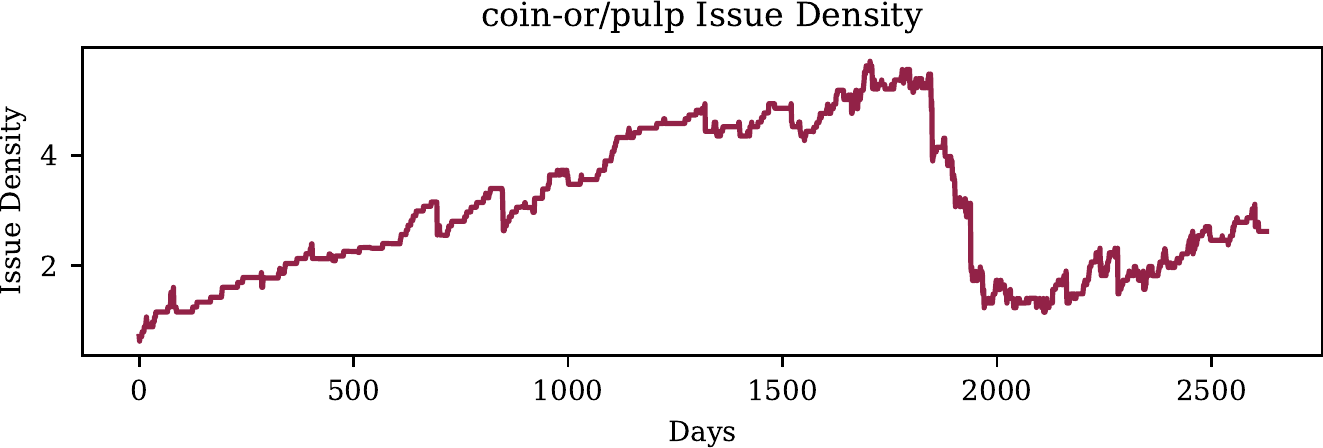}}
    \\
 \subfloat[\label{sub:faker-is}]{%
      \includegraphics[width=0.45\linewidth]{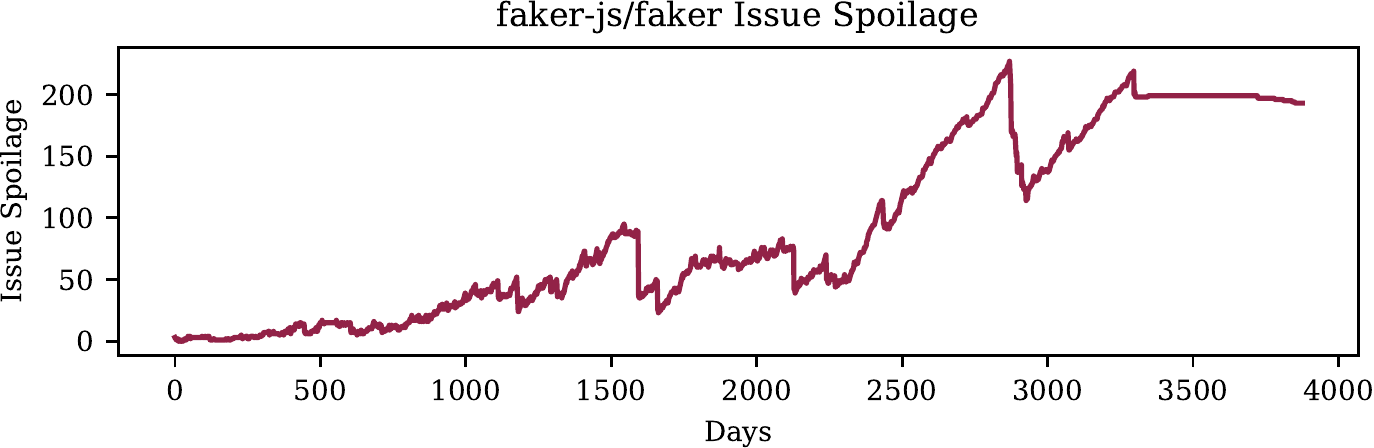}}
    \hfill
 \subfloat[\label{sub:openmalria-is}]{%
      \includegraphics[width=0.45\linewidth]{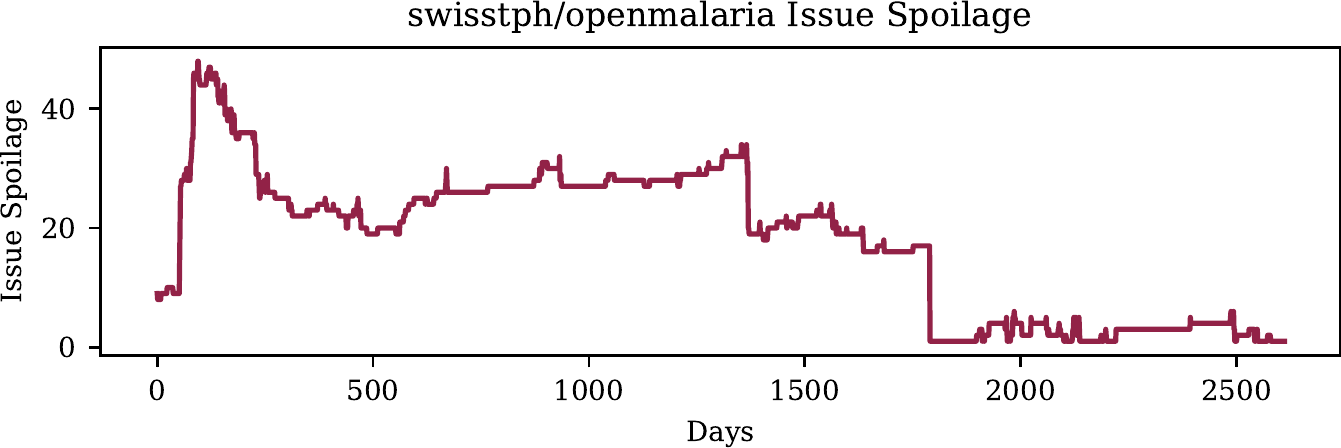}}
    \\
 \subfloat[\label{sub:dronekit-python-p}]{%
      \includegraphics[width=0.45\linewidth]{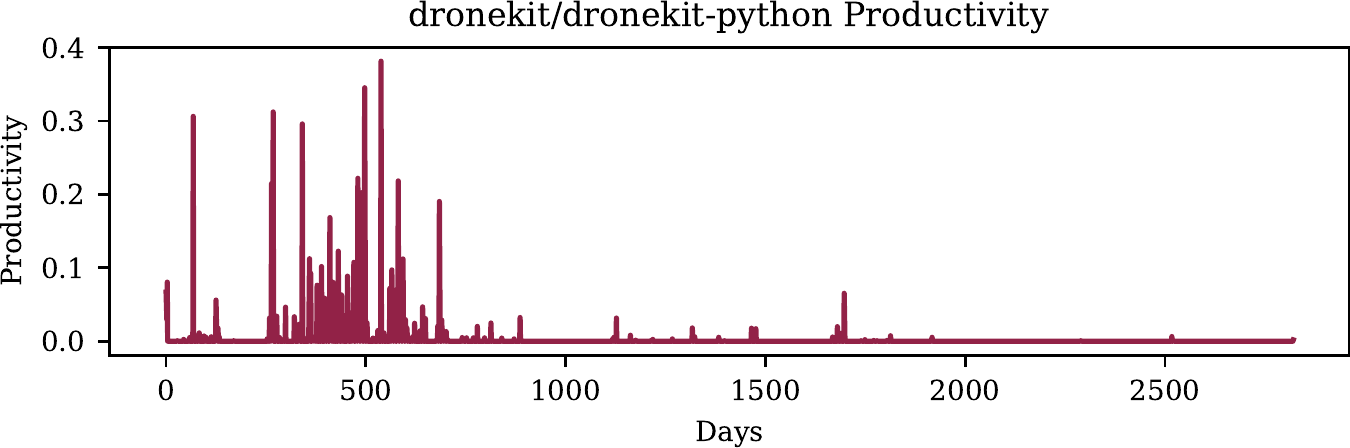}}
    \hfill
 \subfloat[\label{sub:curl-p}]{%
      \includegraphics[width=0.45\linewidth]{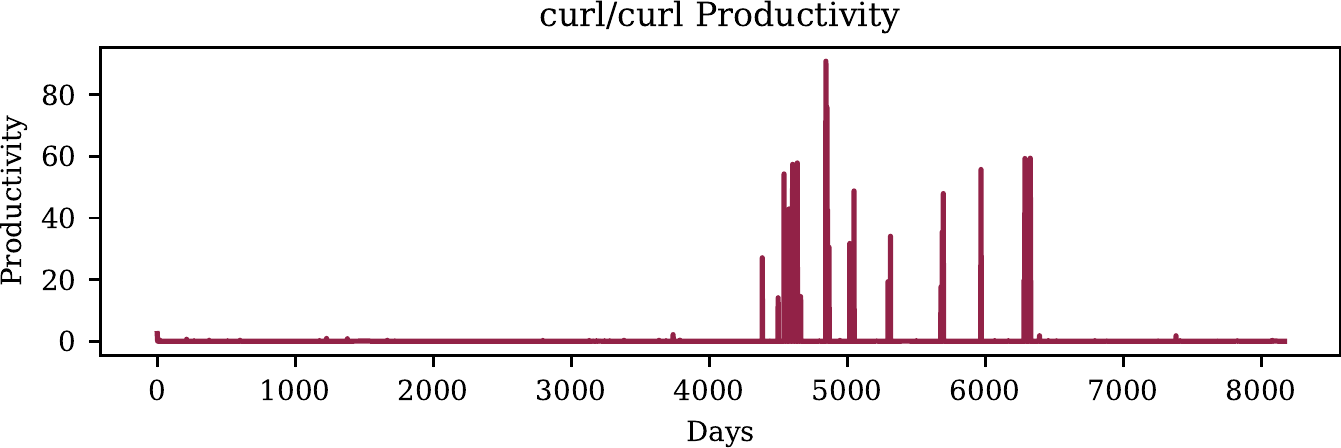}}
    \\
 \subfloat[\label{sub:pulp-bf}]{%
      \includegraphics[width=0.45\linewidth]{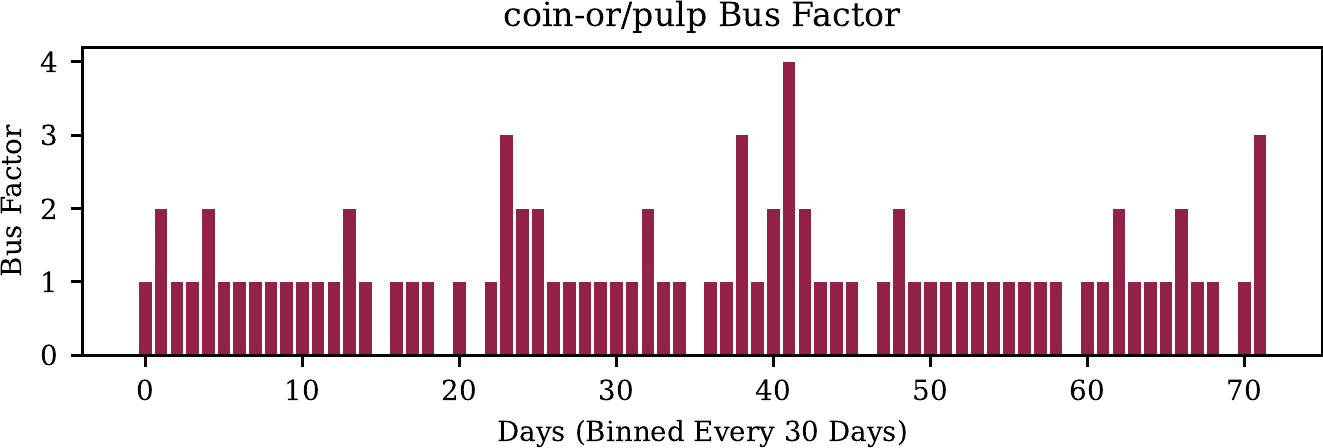}}
    \hfill
 \subfloat[\label{sub:redis-bf}]{%
      \includegraphics[width=0.45\linewidth]{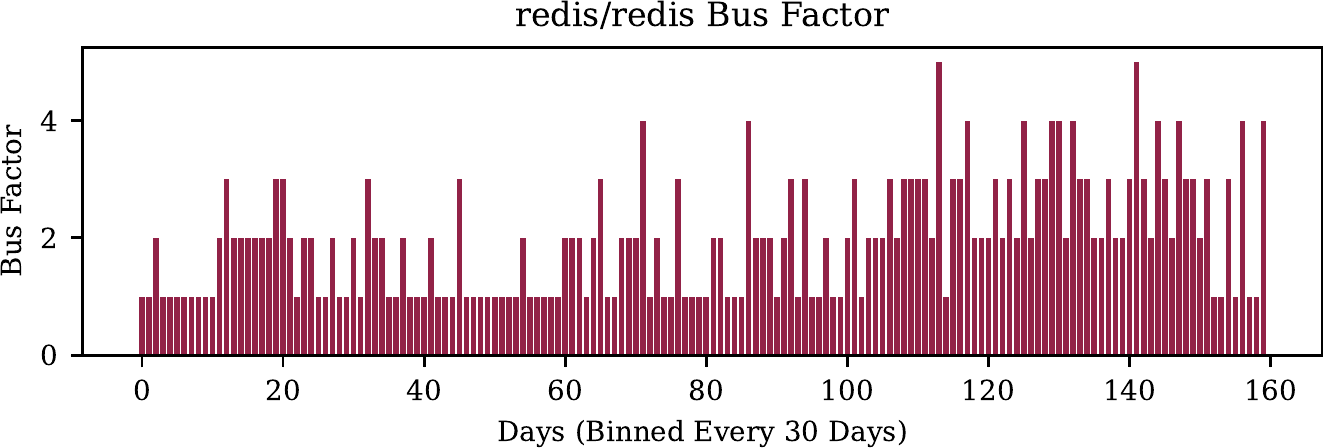}}
    \\

    \caption{
    This figure shows the \tool tool's output for each supported longitudinal derived process metric applied to several sample projects. 
    In the first pair \ref{sub:redis-id}-\ref{sub:pulp-or-id}, \ref{sub:redis-id} sees a significant increase in issue density over its history with some recent improvement. In contrast, \ref{sub:pulp-or-id} saw a similar trend with dramatic improvement in issue density. 
    The second pair \ref{sub:faker-is}-\ref{sub:openmalria-is} shows two projects with contrasting trends in resolving issues.
    The third pair \ref{sub:dronekit-python-p}-\ref{sub:curl-p} shows two projects with contrasting productivity trends. The first had a high level of productivity earlier in its history, while the second is showcases more recent evidence of high productivity.
    The fourth pair \ref{sub:pulp-bf}-\ref{sub:redis-bf} shows two projects with contrasting bus factor.
    The bus factor metric is binned at 30 days to represent the number of core contributors per month.}
    \label{fig:evaluation} 
\end{figure*}
 
\section{Metrics Implemented}
\tool computes two types of software metrics: 
(1) \emph{Direct metrics}, which are measurements of internal attributes of the process, and
(2) \emph{derived metrics}, which are computed metrics from two or more direct metrics.

\subsection{Direct Metrics}

Direct metrics are measurements of a particular attribute of the process involving no other attribute~\cite{fenton_software_2014}.
These measurements are the foundation for the more complex metrics that \tool computes.

\myparagraph{1. Code Size}
\tool measures the size of a repository in terms of the number of source lines of code normalized by 1000 reported as KLOC.
Changes in the KLOC (DKLOC) show the growth (or shrinkage) of a repository over time. 


\myparagraph{2. Developer Count}
\tool measures this metric as the number of unique developers who contribute code to a repository within a time interval.
By measuring developer count, engineering teams can determine the amount of developer support in contributing new code, maintaining existing code, and resolving bugs.

\myparagraph{3. Issue Count}
\tool measures this this metric as the count of the number of open and closed \emph{issues} reported in an issue tracker, including feature requests, tasks, and bug reports, in addition to potential and confirmed defects.
If an online VCS has an issue tracker, this metric also reports the count of open and closed pull requests.

\subsection{Derived Metrics}
Derived metrics are useful for analyzing the interactions between direct metrics~\cite{fenton_software_2014}.
\tool computes derived metrics to analyze and subsequently visualize changes in the development process of a software product. 

\myparagraph{1. Issue Density}
This metric tracks a project's total number of issues normalized by project size.
Because we are interested in open-source repositories on GitHub, we use the more general issue density rather than defect density, which refers only to the ratio of bug count to repository size.
A high issue density, regardless of confirmed defects, could signify an unhealthy repository. 
For example, if there are many feature requests that are never acted upon, then the development team is not implementing the features that users want. 
This would be a possible warning sign for poor customer support and, eventually, would lead to low customer or user satisfaction~\cite{scherkenbach_deming_2011}.

\myparagraph{2. Issue Spoilage}
\label{sec:issue-spoilage}
Issue spoilage is the weighted average age of unresolved issues at a given time in the project timeline.
With further analysis, this metric calculates the age of issues with respect to the project timeline to measure how quickly a project's team resolves issues. 
Issue spoilage can serve as a gauge of customer support and the efficiency of software teams in resolving issues.
For instance, if issue spoilage increases in a time interval, new issues are being created faster than the team can resolve old ones. 
On the other hand, if the issue spoilage drops in a time interval, the team resolves previous issues faster than new ones are created. 

\myparagraph{3. Productivity}
\label{sec:productivity}
Productivity measures the rate at which a development team adds KLOC within a time interval~\cite{fenton_software_2014}. 
Healthy repositories will typically have high productivity.
However, low productivity is not always a sign of a lack of productiveness, as when efficient development teams are refactoring code KLOC may not change significantly.

\myparagraph{4. Bus Factor}
\label{sec:bus-factor}
Bus factor~\cite{cosentino_assessing_2015} is the number of developers on a project team who would have to be hit by a bus to cause the project to fail.
This metric measures the employee turnover risk of a project.
However, as our work focuses on open-source projects, we propose that this is a metric of the development community's interest.
As interest in a project fluctuates, we propose that a longitudinal analysis of this metric is key to assessing interest.



\section{Demonstration}
\label{sec:demonstration}

Figure \ref{fig:evaluation} shows all four process metrics for several repositories over their entire project history. 
We chose projects from the \textsc{RepoReapers/reaper} data set~\cite{munaiah_curating_2017} in pairs that showed contrasting trends in their process metrics to demonstrate possible insights from longitudinal analysis.
We have organized this figure to demonstrate the potential for comparative analysis of process effectiveness, even among projects that have a good score using existing \emph{scorecard} apps.
The addition of process metrics clearly demonstrates that all of these otherwise good projects may benefit from further examining their development process. This examination is especially prudent when it comes to addressing issues (issue spoilage), managing development while addressing issues (issue density), ensuring appropriate resources (bus factor), or managing group priorities to avoid team burnout (productivity).

\section{Planned Studies}
\label{sec:planned-studies}

We intend to expand on the ideas presented in this paper through a few planned studies incorporating \tool.


In the first study, we pose the research question: \emph{How do  engineers use longitudinal process metrics during their development process?}
We hypothesize that basic metrics are used in many open-source projects today, but the use of longitudinal metrics, particularly process metrics, is limited.
To perform this study, we will measure the number of process metrics utilized and survey open-source developers on established projects about why and how they use these metrics in their development process.

In a second study, we pose the research question: \emph{Do longitudinal metrics contribute to selecting dependencies in software composition?}
Based on our survey of tools, we hypothesize that engineers take little consideration of derived longitudinal process metrics but will consider direct longitudinal process metrics as those are more prevalent when selecting dependencies for software development.
To perform this study, we intend to survey the current state of software metrics tooling, and survey open-source engineers about their utilization of longitudinal direct and derived process metrics for dependency selection.

In our third study, we pose the research question: \emph{What role can longitudinal process metrics play in analyzing dependencies in open-source software?}
We hypothesize that many projects are likely to depend on other projects that require process improvement, e.g., a third-party library with a risky bus factor (of one).
To perform this study, we will examine the \emph{dependencies} of well-known projects by using our tools to analyze each of the dependent projects for process-related concerns.






\section{Acknowledgments}


Davis acknowledges support from NSF OAC-2107230.
Thiruvathukal acknowledges support from NSF OAC-2107020 and NSF OAC-1445347.
Davis and Thiruvathukal acknowledge support from the Google TensorFlow Model Garden.

\section{Conclusion}

\tool is an ongoing development effort to understand process effectiveness beyond snapshots of process metrics and support more longitudinal analysis and visualization.
This paper demonstrates working software to compute four process metrics, which represent classical and modern/agile metrics.
We argue for the potential of these tools to support future planned studies by showing their ability to visualize long and short-term trends via simple and intuitive charts.
Future development efforts will include expanding \tool with support for more process metrics, emphasizing comparative visualizations, and expanding the number of data sources.
Future studies will build on this foundation to study the usage of longitudinal metrics in practice, longitudinal metrics in selecting dependencies, and the software supply chain.




\bibliography{refs/adhocbib}
\bibliographystyle{refs/IEEEtran}

%

\end{document}

%% file: settings.tex
\usepackage{amsmath,amssymb,amsfonts}
\usepackage{algorithmic}
\usepackage{graphicx}
\usepackage{textcomp}
\usepackage{xcolor}
\usepackage{booktabs} 
\usepackage{xspace} 
\usepackage[normalem]{ulem}
\usepackage{makecell}
\usepackage{hyperref}

\usepackage{balance} 
\usepackage{caption}
\usepackage{soul}
\usepackage{titlesec}
\usepackage{amsmath}
\usepackage{cleveref}
\usepackage{enumitem}
\usepackage{booktabs}
\usepackage{xcolor}
\usepackage{adjustbox}
\usepackage{tabularx}
\usepackage{todonotes}
\usepackage{paralist}

\crefformat{section}{\S#2#1#3}
\crefname{figure}{Figure}{Figures}
\crefname{appendix}{Appendix}{Appendices}
\crefname{table}{Table}{Tables}
\crefname{algorithm}{Algorithm}{Algorithms}
\crefname{listing}{Listing}{Listings}
\crefname{theorem}{Theorem}{Theorems}
\crefname{thm}{Theorem}{Theorems}
\crefname{lemma}{Lemma}{Lemmata}
\crefname{equation}{Eqt.}{Eqts.}
\crefformat{Grammar}{Grammar #1}

\ifCLASSOPTIONcompsoc
  \usepackage[caption=false,font=normalsize,labelfont=sf,textfont=sf]{subfig}
\else
  \usepackage[caption=false,font=footnotesize]{subfig}
\fi

\newif\ifDEBUG
\DEBUGtrue
\newif\ifANONYMOUS
\ANONYMOUStrue

\ifDEBUG
    \newcommand{\JD}[1]{\textcolor{brown}{[JD: #1]}}
    \newcommand{\WJ}[1]{\textcolor{blue}{[WJ: #1]}}
    \newcommand{\GKT}[1]{\textcolor{brown}{[GKT:#1]}}
    \newcommand{\NMS}[1]{\textcolor{orange}{[NMS: #1]}}
    \newcommand{\RS}[1]{\textcolor{purple}{[RS:#1]}}
    
\else
    \newcommand{\JD}[1]{}
    \newcommand{\WJ}[1]{}
    \newcommand{\GKT}[1]{}
    \newcommand{\RS}[1]{}
    \newcommand{\NMS}[1]{}
    
\fi

\usepackage{etoolbox}
\makeatletter
\patchcmd{\ttlh@hang}{\parindent\z@}{\parindent\z@\leavevmode}{}{}
\patchcmd{\ttlh@hang}{\noindent}{}{}{}
\makeatother

\newcolumntype{R}[2]{%
    >{\adjustbox{angle=#1,lap=\width-(#2)}\bgroup}%
    l%
    <{\egroup}%
}

\newcommand{\myparagraph}[1]{\emph{#1}:}

\newcommand\tool{\textit{PRIME}\xspace}
\IEEEoverridecommandlockouts

%% file: authors.tex

\author{
\IEEEauthorblockN{
  Nicholas Synovic\IEEEauthorrefmark{1},
  Matt Hyatt\IEEEauthorrefmark{1},
  Rohan Sethi\IEEEauthorrefmark{1},
  Sohini Thota\IEEEauthorrefmark{1},
  Shilpika\IEEEauthorrefmark{3},
  Allan J. Miller\IEEEauthorrefmark{1},\\
  Wenxin Jiang\IEEEauthorrefmark{2},
  Emmanuel S. Amobi\IEEEauthorrefmark{1},
  Austin Pinderski\IEEEauthorrefmark{1}\IEEEauthorrefmark{4},
  Konstantin Läufer\IEEEauthorrefmark{1},
  Nicholas J. Hayward\IEEEauthorrefmark{1},\\
  Neil Klingensmith\IEEEauthorrefmark{1},
  James C. Davis\IEEEauthorrefmark{2},
  George K. Thiruvathukal\IEEEauthorrefmark{1}
}


\IEEEauthorblockA{
  \IEEEauthorrefmark{1} Loyola University Chicago, Computer Science Department}

\IEEEauthorblockA{
  \IEEEauthorrefmark{2} Purdue University, Electrical and Computer Engineering Department}

\IEEEauthorblockA{
  \IEEEauthorrefmark{3} University of California at Davis}

  \IEEEauthorrefmark{4} Duke University}